%% file: DDApprox.tex
\documentclass[journal]{IEEEtran}

\usepackage[latin1]{inputenc}
\usepackage{amsmath}
\usepackage{graphicx}
\usepackage{listings}
\usepackage[colorlinks,linkcolor=black,citecolor=black,urlcolor=black]{hyperref}
\usepackage{siunitx}
\usepackage{cite}
\usepackage{fixltx2e}
\usepackage{multirow}
\usepackage{threeparttable}
\usepackage{tabularx}
\usepackage{booktabs}
\usepackage{xcolor}
\usepackage[normalem]{ulem}
\usepackage{float}
\usepackage{caption}
\usepackage{subcaption}

\newif\ifshowToDos

\showToDosfalse

\input{new_commands}

\input{sections/authorblock_ieee.tex}

\begin{document}
\bstctlcite{IEEEexample:BSTcontrol}

\maketitle

\begin{abstract}
  \input{abstract.tex}

\end{abstract}


\section{Introduction}
\label{sec:intro}
\input{sections/intro}

\section{Results}
\label{sec:results}
\input{sections/results}

\section{Discussion}
\label{sec:discussion}
\input{sections/discussion}

\section{Methods}
\label{sec:methods}
\input{sections/methods}

\input{sections/fid}

\bibliographystyle{IEEEtran}
\bibliography{DDApprox}

\clearpage

\end{document}

%% file: new_commands.tex
\ifshowToDos
  \usepackage{todonotes}
  \paperwidth=\dimexpr \paperwidth + 8cm\relax
  \oddsidemargin=\dimexpr\oddsidemargin + 4cm\relax
  \evensidemargin=\dimexpr\evensidemargin + 4cm\relax
  \marginparwidth=\dimexpr \marginparwidth + 4cm\relax
\else
  \usepackage[disable]{todonotes}
\fi

\definecolor{gray}{gray}{0.2}



%% file: sections/authorblock_ieee.tex
\title{Node Replacement based Approximate Quantum Simulation with Decision Diagrams}

\author{Yexin Yan, Stefan Hillmich, Robert Wille, Christian Mayr
\thanks{Manuscript received \today. The research leading to these results has received funding from the European Union (EU) Seventh Framework Programme (FP7) under grant agreement No. 604102, the EU's Horizon 2020 research and innovation programme under grant agreements No. 720270 and 785907 (Human Brain Project, HBP), and the German Federal Ministry of Education and Research (BMBF) and the free state of Saxony within the ScaDS.AI center of excellence for AI research. Additionally, it has received funding from the European Research Council (ERC) under the European Union's Horizon 2020 research and innovation programme (grant agreement No. 101001318). It is part of the Munich Quantum Valley, which is supported by the Bavarian state government with funds from the Hightech Agenda Bayern Plus and was partially supported by the BMK, BMDW, and the State of Upper Austria in the frame of the COMET program (managed by the FFG). }
\thanks{Y. Yan and C. Mayr are with Technische Universit\"at Dresden, Germany and ScaDS.AI Dresden/Leipzig, Germany (e-mail: yexin.yan@tu-dresden.de). }
\thanks{S. Hillmich is with Software Competence Centrum Hagenberg (SCCH), Austria.}
\thanks{R. Wille is with Technische Universit\"at M\"unchen, Germany and Software Competence Centrum Hagenberg (SCCH), Austria.}
}

%% file: abstract.tex
Simulating a quantum circuit with a classical computer requires exponentially growing resources. Decision diagrams exploit the redundancies in quantum circuit representation to efficiently represent and simulate quantum circuits. But for complicated quantum circuits like the quantum supremacy benchmark, there is almost no redundancy to exploit. Therefore, it often makes sense to do a trade-off between simulation accuracy and memory requirement. Previous work on approximate simulation with decision diagrams exploits this trade-off by removing less important nodes. In this work, instead of \textit{removing} these nodes, we try to find similar nodes to \textit{replace} them, effectively slowing down the fidelity loss when reducing the memory. In addition, we adopt Locality Sensitive Hashing (LSH) to drastically reduce the computational complexity for searching for replacement nodes. Our new approach achieves a better memory-accuracy trade-off for representing a quantum circuit with decision diagrams with minimal run time overhead. Notably, our approach shows good scaling properties when increasing the circuit size and depth. For the first time, a strong better-than-linear trade-off between memory and fidelity is demonstrated for a decision diagram based quantum simulation when representing the quantum supremacy benchmark circuits at high circuit depths, showing the potential of drastically reducing the resource requirement for approximate simulation of the quantum supremacy benchmarks on a classical computer.

%% file: sections/intro.tex
Quantum computing is an emerging computational paradigm that promises to provide computational power far beyond classical computers\cite{grover}\cite{shor}\cite{Nielsen_Chuang_2010}. 'Quantum Supremacy' is claimed to be achieved, if a quantum computer performs a task that is impossible for a classical computer to perform within reasonable time and energy \cite{Harrow2017}\cite{Boixo2018}. The quantum supremacy benchmark (also known as random quantum circuit (RQC)) \cite{Boixo2018} was proposed as one such computational task to compare the computational power of a quantum computer and a classical computer. Therefore, simulating the quantum supremacy benchmark on a classical computer more efficiently would narrow down the gap between quantum computers and classical computers. At the same time, simulating the supremacy benchmark on a classical computer also provides verification for the quantum hardware performing the same task \cite{qflex}. 

Different approaches have been proposed for simulating quantum circuits. Tensor network (TN)-based simulators \cite{markov2008}\cite{qflex} can efficiently represent a quantum circuit and calculate a certain quantum state. But calculating the complete state vector still requires exponential time. On the other hand, the decision diagram (DD)-based approach \cite{niemann2016}\cite{zulehner2019} can represent the full state vector, but constructing and representing the quantum circuit with a DD might still require exponentially growing resources in the worst case. 

While exact simulation of the quantum supremacy benchmark almost always requires exponentially growing computational and memory resources, approximate simulation might offer a path to simulate larger and deeper circuits with acceptable accuracy and resources. 

The goal of approximate quantum simulation is to improve the trade-off between memory and accuracy, i.e. to simulate a quantum circuit with higher accuracy and less memory. Additionally, it would be desirable that the approximation algorithm doesn't take too much run time. 

For TN-based approaches, linear trade-off between memory and accuracy when simulating the quantum supremacy circuit has been achieved by adopting the Schr\"odinger-Feynman algorithm and selectively simulating a fraction of the paths \cite{boixo2018b}\cite{qflex}. 

For DD-based approaches, better-than-linear trade-off between memory and accuracy has been achieved by removing nodes with lower contribution \cite{hillmichApprox2022}. But as shown in this work, with the node removal approach, the trade-off becomes more linear when increasing the circuit depth, due to further randomization of the state vector. 

In this work, we focus on reducing the memory requirement of the DD of the quantum supremacy circuit after the DD is constructed. To this end, we propose \textit{replacing} the nodes with similar nodes instead of \textit{removing} them. Additionally, we adopt Locality Sensitive Hashing (LSH) to drastically reduce the computational complexity for finding similar nodes for replacement. 

We demonstrate in this work:

\begin{enumerate}
    \item substantial improvement of memory-accuracy trade-off compared to the previous work \cite{hillmichApprox2022}
    \item 4 node replacement strategies suitable for different fidelity regions
    \item better memory-accuracy trade-off when scaling up the circuit size
    \item an LSH-based algorithm for efficiently searching for similar nodes for replacement
    \item for the first time, a strong better-than-linear trade-off between memory and accuracy for the quantum supremacy benchmark at high circuit depths. 
\end{enumerate}

This paper is organized as follows. In "\nameref{sec:results}" we show the improvement of our approach compared to the previous work, the 4 node replacement strategies, LSH-based node search and scaling properties. In "\nameref{sec:discussion}" we summarize the results achieved in this work and discuss about limitations and future work. In "\nameref{sec:methods}" we present the details of our approach, including quantifying the fidelity loss when replacing nodes, the 4 node replacement strategies, reasons for choosing the distance metric for LSH, and the implementation of hierarchical Super-Bit LSH.

%% file: sections/results.tex
The simulation results are obtained with the quantum supremacy benchmarks \cite{Boixo2018}, since they represent the worst case where little redundancy can be exploited. But our approach is generally applicable to other quantum algorithms. First, we show the results of using LSH for node replacement. Then we show the results of node replacement strategies. The results for the node replacement strategies are obtained using LSH. 

For simplicity, the different node replacement strategies are summarized as "NxX", meaning replacing N levels for X times (see "\nameref{sec:methods}"). Under this convention, single level node replacement is shown as "1x1", multiple level node replacement is shown as "Nx1", independent node replacement is shown as "1xX" and independent multiple level node replacement is shown as "NxX".

\subsection{Locality Sensitive Hashing}\label{subsec:res_lsh}

LSH results in drastically reduced program run time for searching for replacement nodes, while maintaining similar fidelity compared to the exhaustive approach (see "\nameref{subsec:lsh_idea}"). The node replacement strategy 1x1 (see "\nameref{subsec:single_level}") is chosen to demonstrate the effect of this approach. 

\begin{figure}
    \centering\includegraphics[width=1\linewidth]{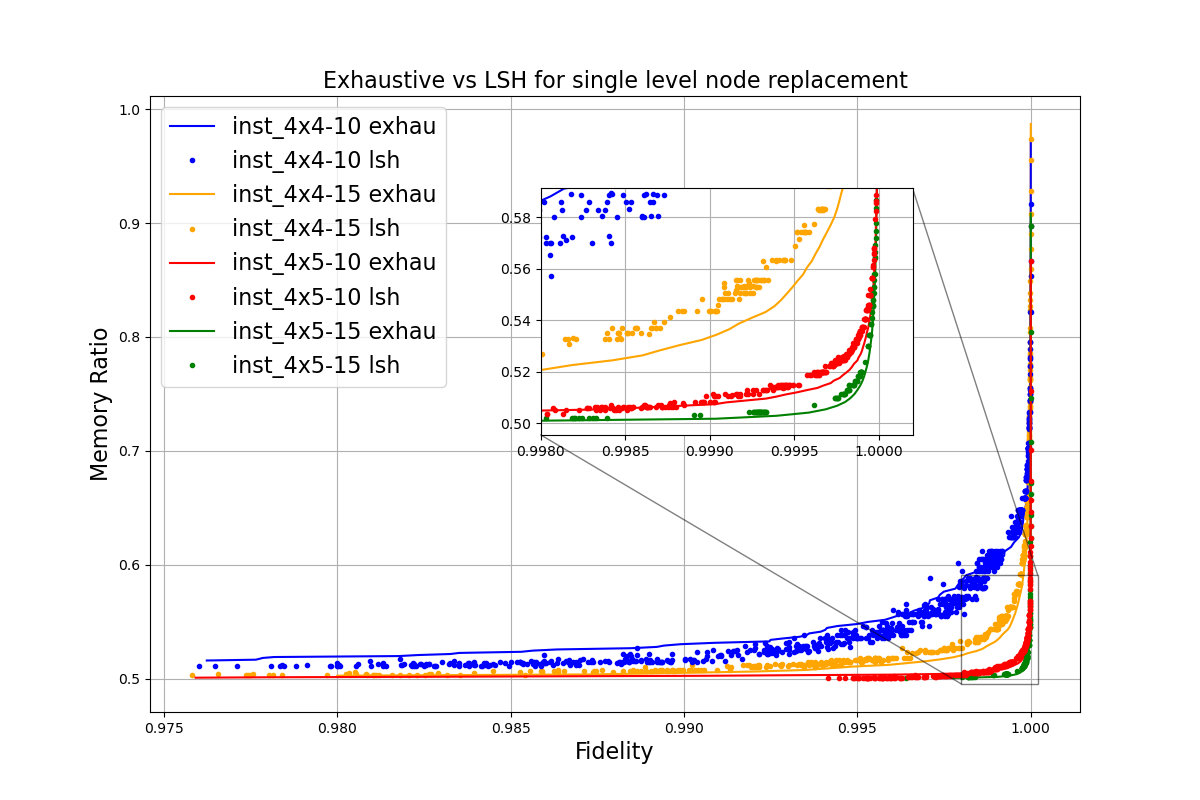}
    \caption{Memory ratio vs. fidelity for exhaustive search (solid lines) and LSH (dots). Quantum supremacy circuits with 4x4 and 4x5 qubits and depths of 10 and 15 are shown (benchmarks inst4x4\_10\_0, inst4x4\_15\_0, inst4x5\_10\_0, inst4x5\_15\_0). LSH provides a good approximation of the exhaustive approach.}
    \label{fig:lsh_memratio}
\end{figure}

In Figure \ref{fig:lsh_memratio}, we show LSH provides a good approximation of the exhaustive approach by comparing LSH with exhaustive approach on 4 benchmark circuits: inst\_4x4\_10\_0, inst\_4x4\_15\_0, inst\_4x5\_10\_0 and inst\_4x5\_15\_0. The root mean square error (RMSE) of memory ratio between LSH and exhaustive is summarized in Table \ref{tab:lsh_runtime}. The error caused by LSH is around 1\% for inst\_4x4\_10\_0 and inst\_4x4\_15\_0, and under 1\% for inst\_4x5\_10\_0 and inst\_4x5\_15\_0.

In addition, in Table \ref{tab:lsh_runtime}, we show the run time speed-up for searching for replacement nodes, with the same 4 benchmark circuits as above. The run time is obtained with 50\% replaced nodes and 50\% replacement nodes in the lowest level of the DD. The number of nodes in the lowest level is also shown in the table, to indicate the computational complexity of different benchmarks. The simulations were run on AMD EPYC CPU 7352. The data shown is the average run time of 5 runs. From the results, we observe a speed-up from 79.3 times to 3064.1 times. We assume this speed-up factor would continue to increase as we simulate larger circuits. But simulating larger circuits were not possible with available compute resources.

\begin{table}[h]
    \centering
    \begin{tabular}{|c|c|c|c|c|}
    \hline
        Benchmark & 4x4\_10\_0 & 4x4\_15\_0 & 4x5\_10\_0 & 4x5\_15\_0\\
        \hline
        nr. of qubits & 16 & 16 & 20 & 20\\
        nr. of nodes & 31731 & 32768 & 524287 & 524287 \\
        \hline
        RMSE &  0.01286 &  0.01163 & 0.008532 &  0.004978 \\
        \hline
        exhau. run time & 4.78 s & 5.25 s & 5719.19 s & 6195.03 s \\
        LSH run time & 0.0603 s & 0.0599 s & 2.2394 s & 2.0218 s \\
        speed-up   & \textbf{79.3X} & \textbf{87.6X} & \textbf{2553.9X} & \textbf{3064.1X}\\
        \hline
        
    \end{tabular}
    \caption{RMSE between LSH and exhaustive approach, and speed-up of run time for searching for replacement nodes evaluated on quantum supremacy benchmarks with 4x4 and 4x5 qubits with depth 10 and 15.}
    \label{tab:lsh_runtime}
\end{table}

\subsection{Performance of Single Level Node Replacement}\label{subsec:res_single_level}

To demonstrate the performance of single level node replacement(1x1, see "\nameref{subsec:single_level}"), the fidelity and relative memory are tested with the quantum supremacy benchmark (inst4x4\_10\_0 and inst4x5\_10\_0) and compared to previous work \cite{hillmichApprox2022}. As shown in Fig. \ref{fig:independentlevel}, when reducing the fidelity, the memory first drops fast, then it saturates. The relative memory and the fidelity of a few data points in Fig. \ref{fig:independentlevel} are listed in Table \ref{tab:singlelevel1} and \ref{tab:singlelevel2}.

\begin{table}[h]
    \centering
    \begin{tabular}{|c|c|c|c|c|}
    \hline
        Fidelity & 99.99\% & 99.9\% & 99.0\% & 97.0\%\\
        \hline
        Memory (previous)   & 99.7\% & 97.8\% & 89.8\% & 80.0\%\\
        \hline
        Memory (new)   & 70.0\% & 60.4\% & 51.7\% & 51.1\%\\
        \hline
        
    \end{tabular}
    \caption{Comparison of relative memory for single level node replacement (this work) and previous work for the inst4x4\_10\_0 benchmark.}
    \label{tab:singlelevel1}
\end{table}
\begin{table}[h]
    \centering
    \begin{tabular}{|c|c|c|c|c|}
    \hline
        Fidelity  & 99.9997\% & 99.9993\% & 99.99\% & 99.9\%\\
        \hline
        Memory (previous)  & 99.993\% & 99.98\% & 99.7\% & 97.4\% \\
        \hline
        Memory (new)    & 70.0\% & 60.0\% & 53.6\% & 50.7\%\\
        \hline
        
    \end{tabular}
    \caption{Comparison of relative memory for single level node replacement (this work) and previous work for the inst4x5\_10\_0 benchmark.}
    \label{tab:singlelevel2}
\end{table}

As shown in Table \ref{tab:singlelevel1} and \ref{tab:singlelevel2}, single level node replacement achieves 99.9\% (99.9993\%) fidelity with 60.4\% (60.0\%) memory in the inst4x4\_10\_0 (inst4x5\_10\_0) benchmark, which is better than the previous Sota, but the memory reduction saturates at around 50\%. The reason is that only nodes on the lowest level are replaced, which is at most 50\% of all nodes. To further improve memory reduction in the lower fidelity region, it would be necessary to replace nodes on multiple levels.

\subsection{Performance of Multiple Level Node Replacement}\label{subsec:res_multiple_level}

As shown in Fig. \ref{fig:independentlevel}, compared to single level node replacement, by replacing multiple levels (2x1 and 3x1, see "\nameref{subsec:multiple_level}"), the lower bound for memory reduction is decreased, because now more nodes can be considered for replacement. In addition, for the same reason, replacing 3 levels further decreases the lower bound compared to replacing 2 levels. But in the upper right region, where both memory and fidelity are high, replacing multiple levels results in more fidelity loss for the same memory. In addition, replacing 3 levels causes more fidelity loss than 2 levels. The reason is, when replacing multiple levels, all child nodes of the replaced node are "bundled". And when replacing more levels, e.g. 3 levels, more child nodes are bundled than replacing less levels, e.g. 2 levels. But when further decreasing memory and fidelity, the lines cross, because now the advantage of the smaller lower bound for memory dominates. 

\subsection{Performance of Independent Node Replacement}\label{subsec:res_iterative}

 \begin{figure}
    
    \begin{subfigure}{1\linewidth}
    \centering\includegraphics[width=1\linewidth]{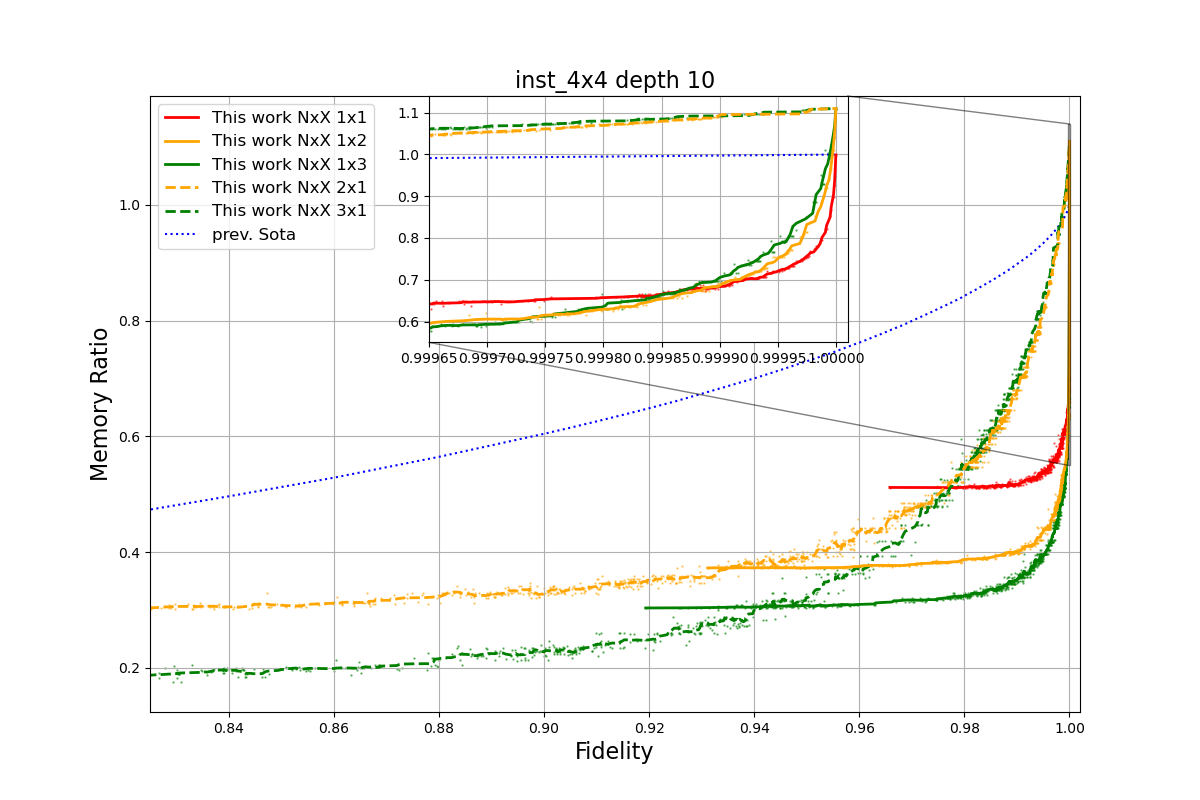}
    \caption{inst4x4 with depth 10.}
    \label{fig:indmult1}
    \end{subfigure}
    \begin{subfigure}{1\linewidth}
    \centering\includegraphics[width=1\linewidth]{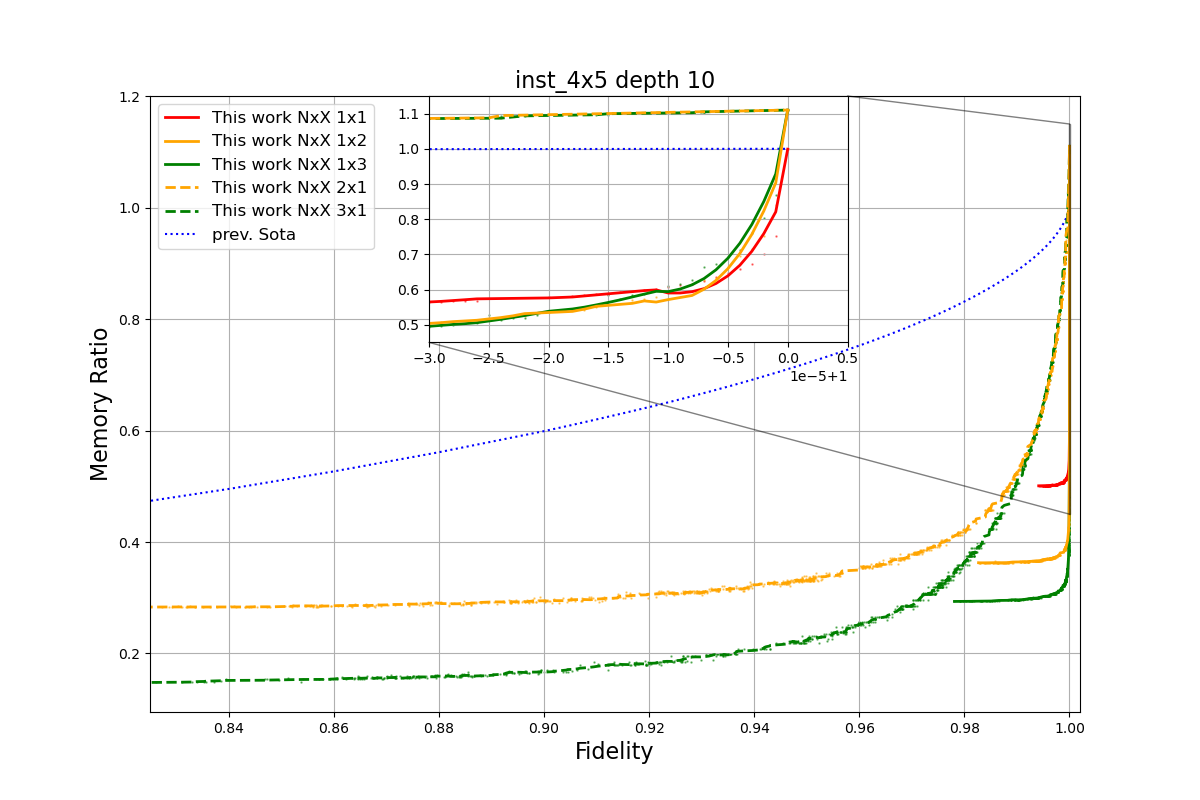}
    \caption{inst4x5 with depth 10.}
    \label{fig:indmult2}
    \end{subfigure}
    \caption{Single level node replacement (1x1), multiple level node replacement (2x1 and 3x1) and independent node replacement (1x2 and 1x3) in comparison to the previous Sota (blue dotted line) \cite{hillmichApprox2022} tested on the quantum supremacy benchmarks (inst4x4\_10\_0 and inst4x5\_10\_0). Single level node replacement (red solid line) achieves the fastest memory reduction in the high fidelity range, but the memory reduction saturates at around 50\%. The yellow dashed line ("2x1") shows 2 level node replacement and the green dashed line ("3x1") shows 3 level node replacement. When more levels are replaced, the memory reduction in the high fidelity range slows down, but in the low fidelity region the curve saturates at a lower point. The yellow solid line ("1x2") shows 2 independent level node replacement and the green solid line ("1x3") shows 3 independent level node replacement. The advantage of independent replacement on each level results in faster memory reduction in the high fidelity region compared to multiple level node replacement. When more levels are involved in independent node replacement, the memory reduction in high fidelity region slows down (see insert). Due to the memory overhead of virtual edges, the curves saturate at a higher point compared to multiple level node replacement. For 1x2, 1x3, 2x1 and 3x1, the memory ratio for fidelity 1.0 is 1.11, due to the memory overhead of the node. For the simplest case 1x1, the memory overhead is not necessary. For the results of node replacement (this work), the dots are the simulation results. The curves are obtained by applying a Savitzky-Golay filter with window length 21 and poly order 3 to the dots.}
    \label{fig:independentlevel}
\end{figure}

Compared to multiple level node replacement, allowing nodes to be replaced independently on multiple levels (Fig. \ref{fig:independentlevel}, 1x2, 1x3) slows down the fidelity loss in the high fidelity region, at the cost of memory overhead of virtual edges (see "\nameref{subsec:iterative}"). As a result, the memory-accuracy trade-off is between single level node replacement and multiple level node replacement. Compared to single level node replacement, the memory reduction saturates at a lower point because replacement is done on more levels. In addition, memory reduction saturates at a lower point with 3 level replacement than with 2 level replacement, due to replacement on more levels. But in the high fidelity region, the fidelity drops faster due to the memory overhead. In addition, fidelity drops faster with 3 level replacement than with 2 level replacement due to more memory overhead. Compared to multiple level node replacement, the fidelity drop is slower in the high fidelity region due to independent replacement on each level, but the memory reduction saturates at a higher point due to the memory overhead.

\begin{figure}
    
    \begin{subfigure}{1\linewidth}
    \centering\includegraphics[width=1\linewidth]{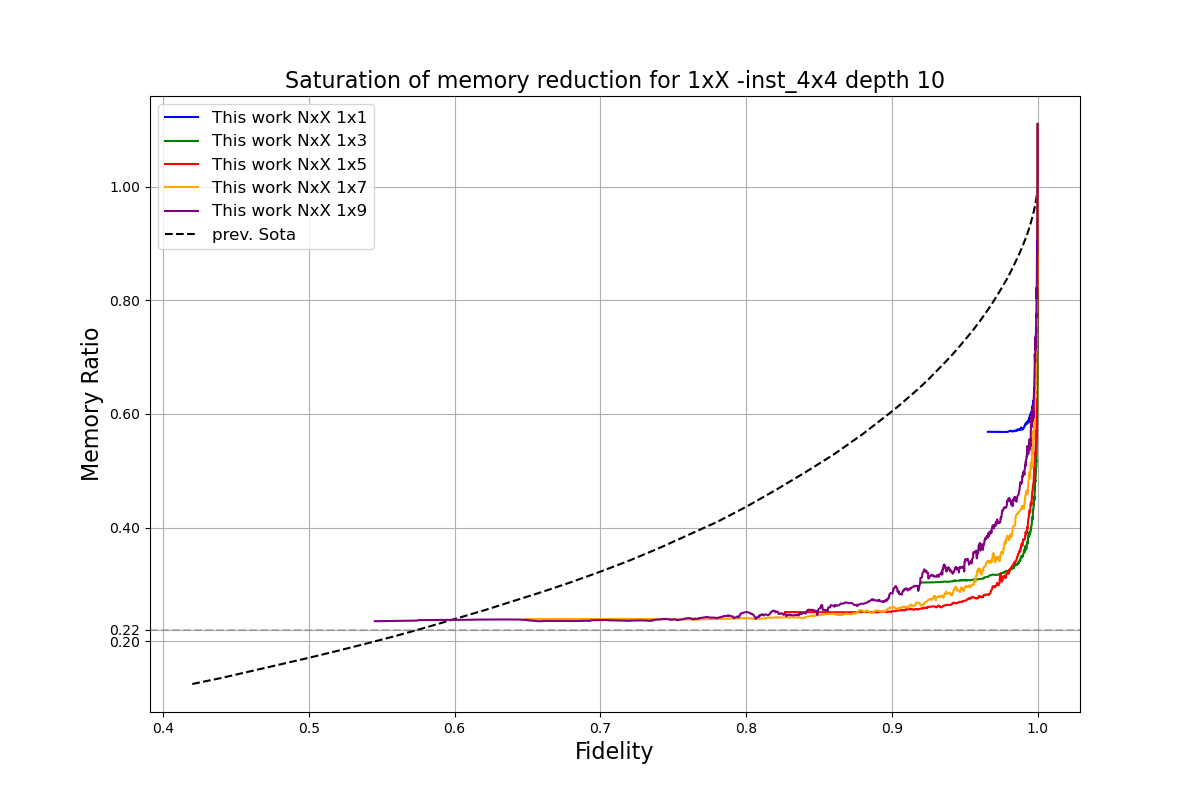}
    \caption{inst4x4 with depth 10.}
    \label{fig:indmult1}
    \end{subfigure}
    \begin{subfigure}{1\linewidth}
    \centering\includegraphics[width=1\linewidth]{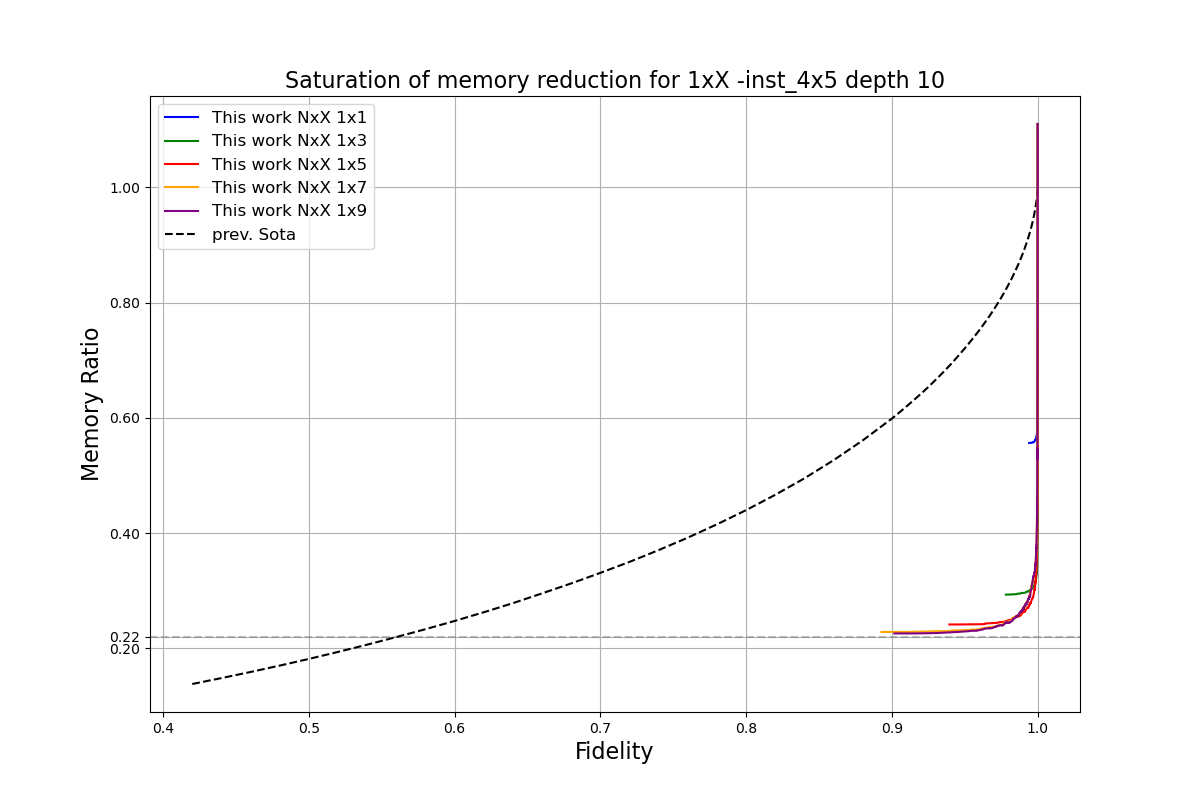}
    \caption{inst4x5 with depth 10.}
    \label{fig:indmult2}
    \end{subfigure}
    \caption{The saturation point of memory reduction converges to 22\% in the case of 1xX due to memory overhead.}
    \label{fig:memreduction1xX}
\end{figure}

As shown in Fig. \ref{fig:memreduction1xX}, when increasing the number of levels for independent node replacement, the memory reduction saturates at a lower point, because more nodes are replaced. However, this reduction of saturation point converges to 22\% in the case of 1xX due to memory overhead (see "\nameref{subsec:iterative}"). 

\subsection{Performance of Independent Multiple Level Node Replacement}\label{subsec:res_multi_iterative}

\begin{figure}
    
    \begin{subfigure}{1\linewidth}
    \centering\includegraphics[width=1\linewidth]{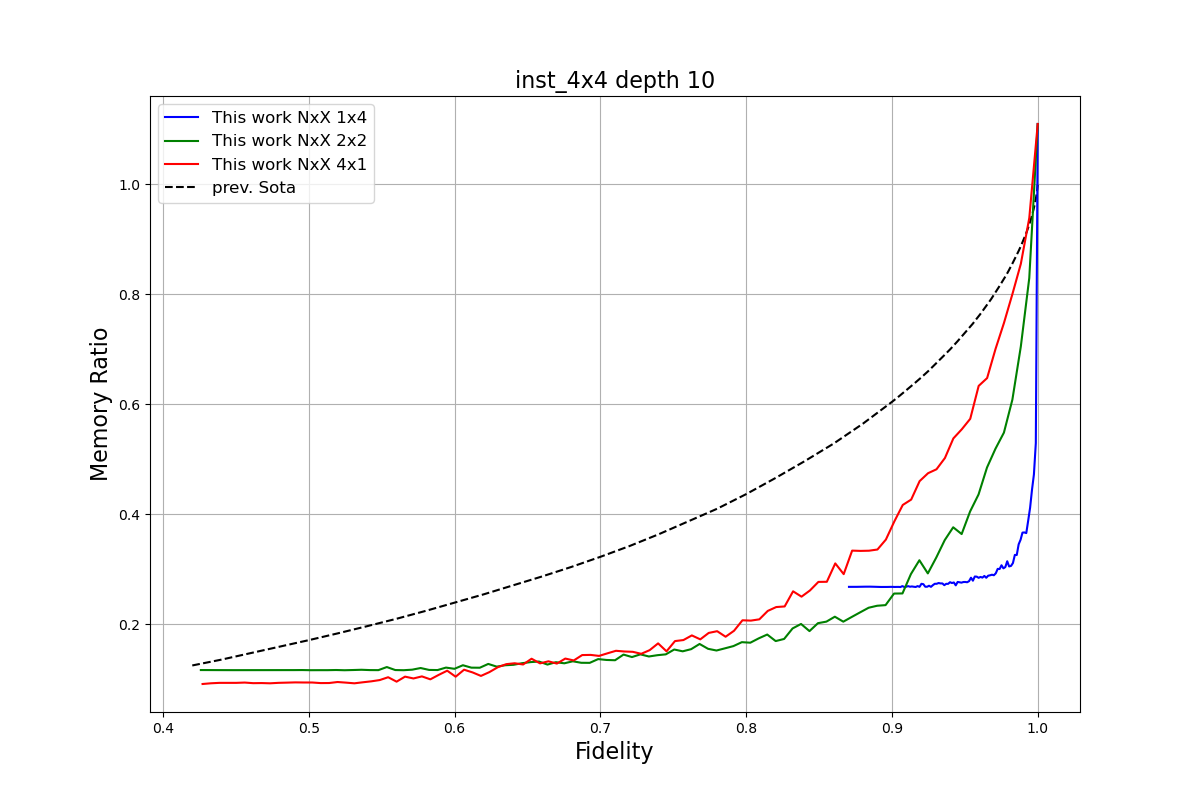}
    \caption{inst4x4 with depth 10.}
    \label{fig:indmult1}
    \end{subfigure}
    \begin{subfigure}{1\linewidth}
    \centering\includegraphics[width=1\linewidth]{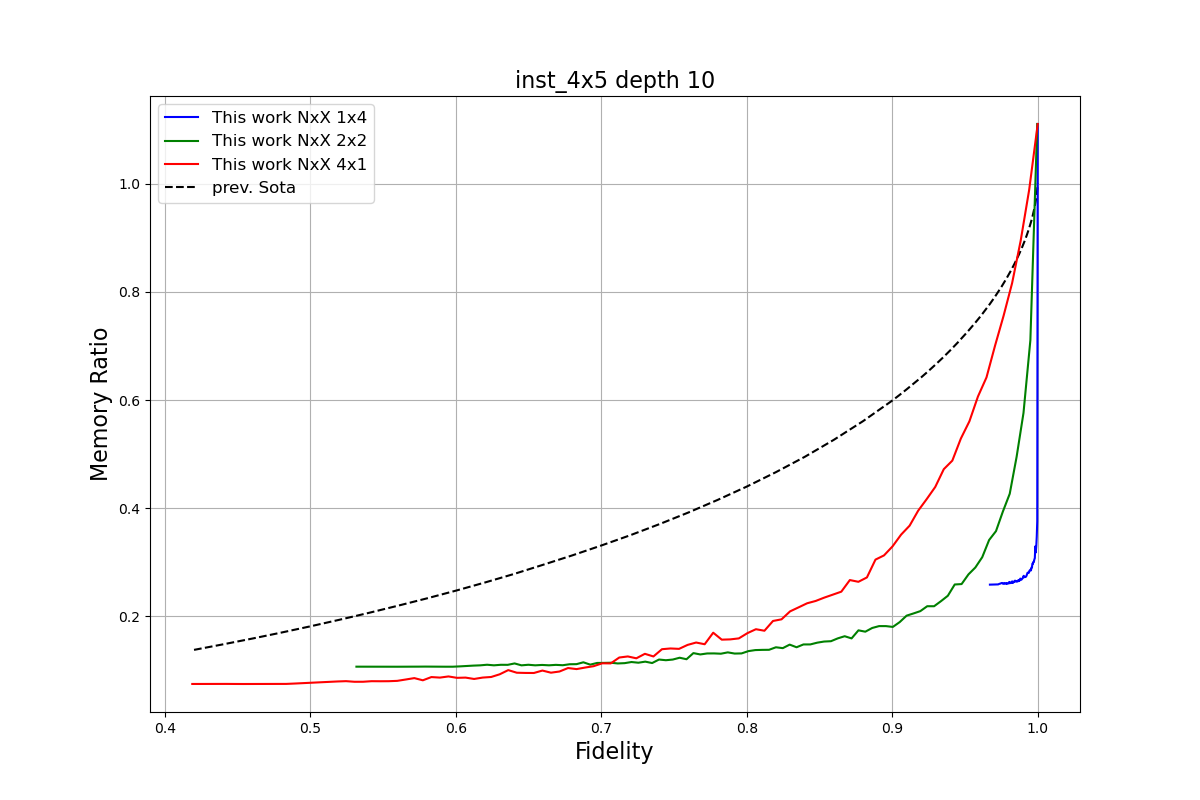}
    \caption{inst4x5 with depth 10.}
    \label{fig:indmult2}
    \end{subfigure}
    \caption{Independent multiple level node replacement (green lines), in comparison to the previous SOTA (dashed line) \cite{hillmichApprox2022}, multiple level node replacement (red line) and independent node replacement (blue line), tested on the quantum supremacy benchmark (inst4x4\_10\_0 and inst4x5\_10\_0). The node replacement is done once every $N$ levels for $X$ times ($N$x$X$). The combination of independent node replacement and multiple level node replacement provides an intermediate solution between both strategies and creates a better memory reduction in the intermediate fidelity regions.}
    \label{fig:indmult}
\end{figure}

As shown in Fig. \ref{fig:indmult}, independent multiple level node replacement (see "\nameref{subsec:multi_iterative}") combines multiple level node replacement and independent node replacement and provides a solution between both strategies: the fidelity loss becomes slower than multiple level node replacement due to the independent replacement, and the memory bound becomes lower than independent node replacement due to less memory overhead. Overall, the hybrid approach provides a better solution where the memory reduction of independent node replacement in the higher fidelity region already slows down but the advantage of the lower memory bound of multiple level node replacement in the lower fidelity region is not yet dominant.

\begin{figure}
    
    \begin{subfigure}{1\linewidth}
    \centering\includegraphics[width=1\linewidth]{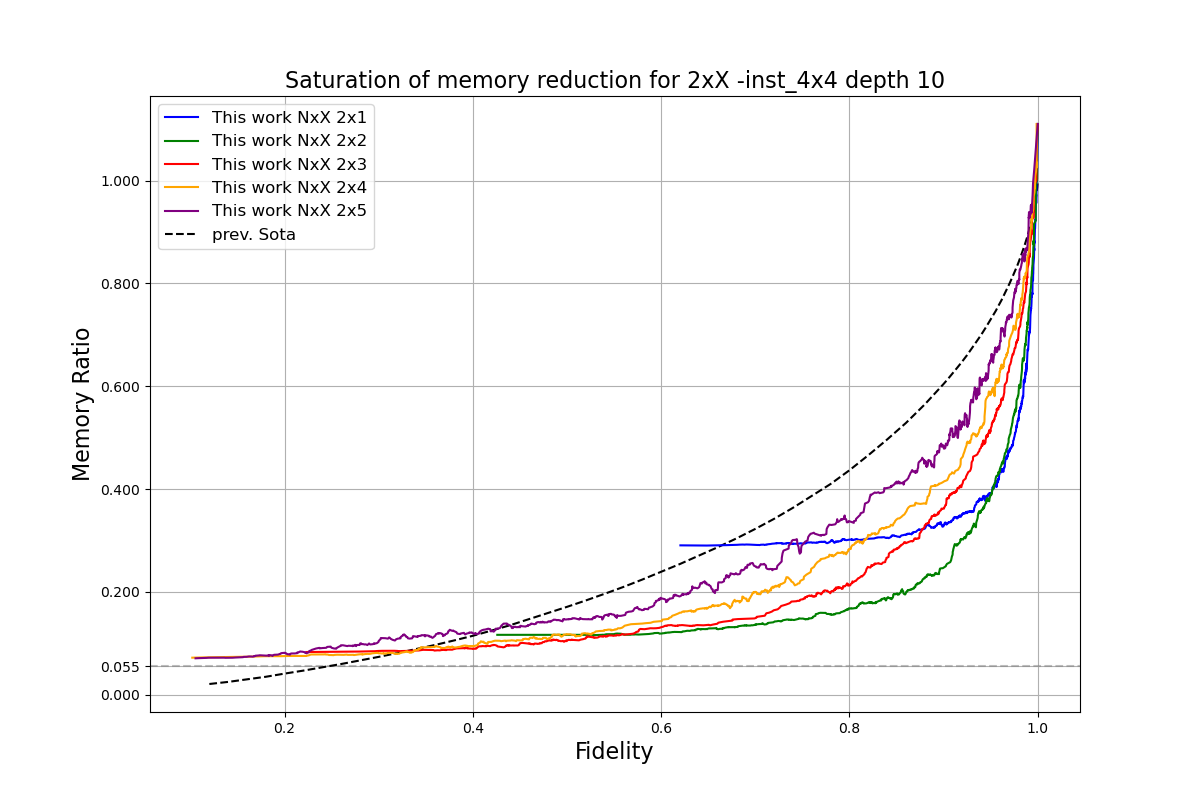}
    \caption{inst4x4 with depth 10.}
    \label{fig:indmult1}
    \end{subfigure}
    \begin{subfigure}{1\linewidth}
    \centering\includegraphics[width=1\linewidth]{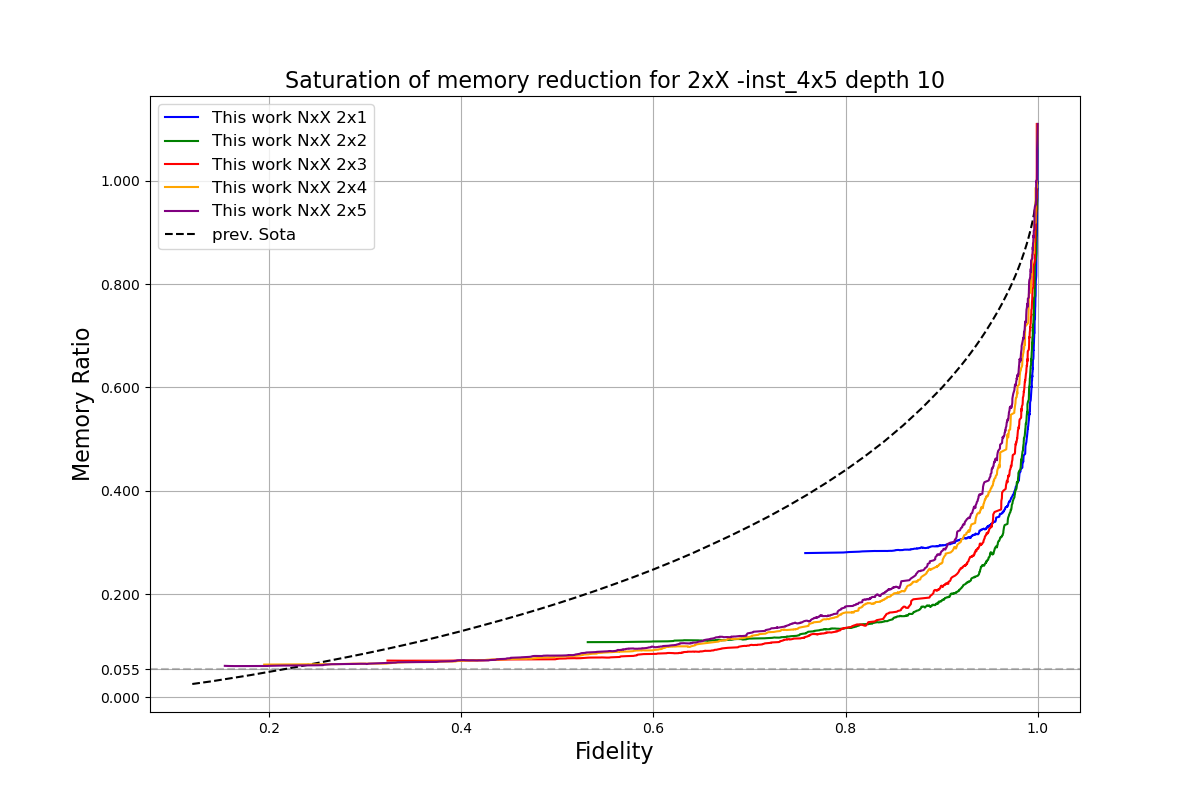}
    \caption{inst4x5 with depth 10.}
    \label{fig:indmult2}
    \end{subfigure}
    \caption{The saturation point of memory reduction converges to 5.5\% in the case of 2xX due to memory overhead.}
    \label{fig:memreduction2xX}
\end{figure}

As shown in Fig. \ref{fig:memreduction2xX}, similar to Fig. \ref{fig:memreduction1xX}, when increasing the number of levels for independent node replacement, the memory reduction saturates at a lower point, and this reduction of saturation point converges to 5.5\% in the case of 2xX (see "\nameref{subsec:multi_iterative}").

\subsection{Scaling up Circuit Size}\label{subsec:res_size}

\begin{figure}
    
    \begin{subfigure}{1\linewidth}
    \centering\includegraphics[width=1\linewidth]{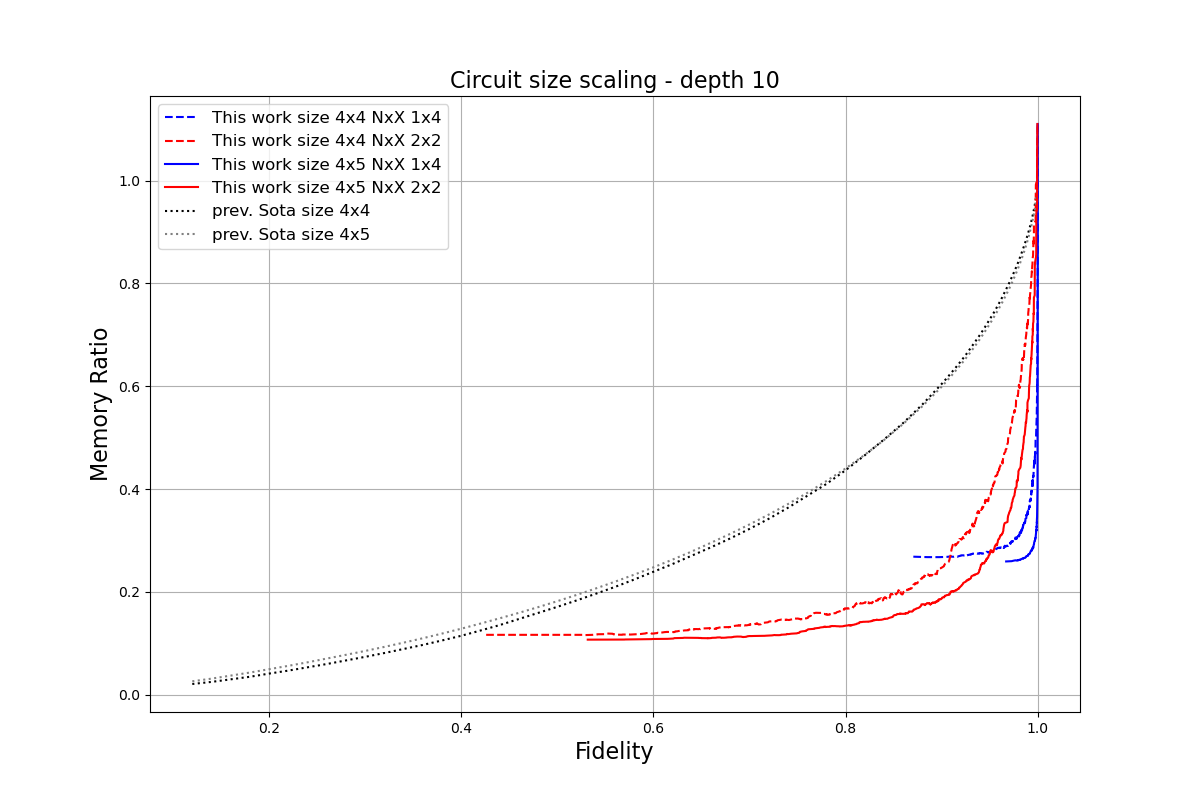}
    \caption{depth 10}
    \label{fig:lsh_fidmem_4x4}
    \end{subfigure}
    \begin{subfigure}{1\linewidth}
    \centering\includegraphics[width=1\linewidth]{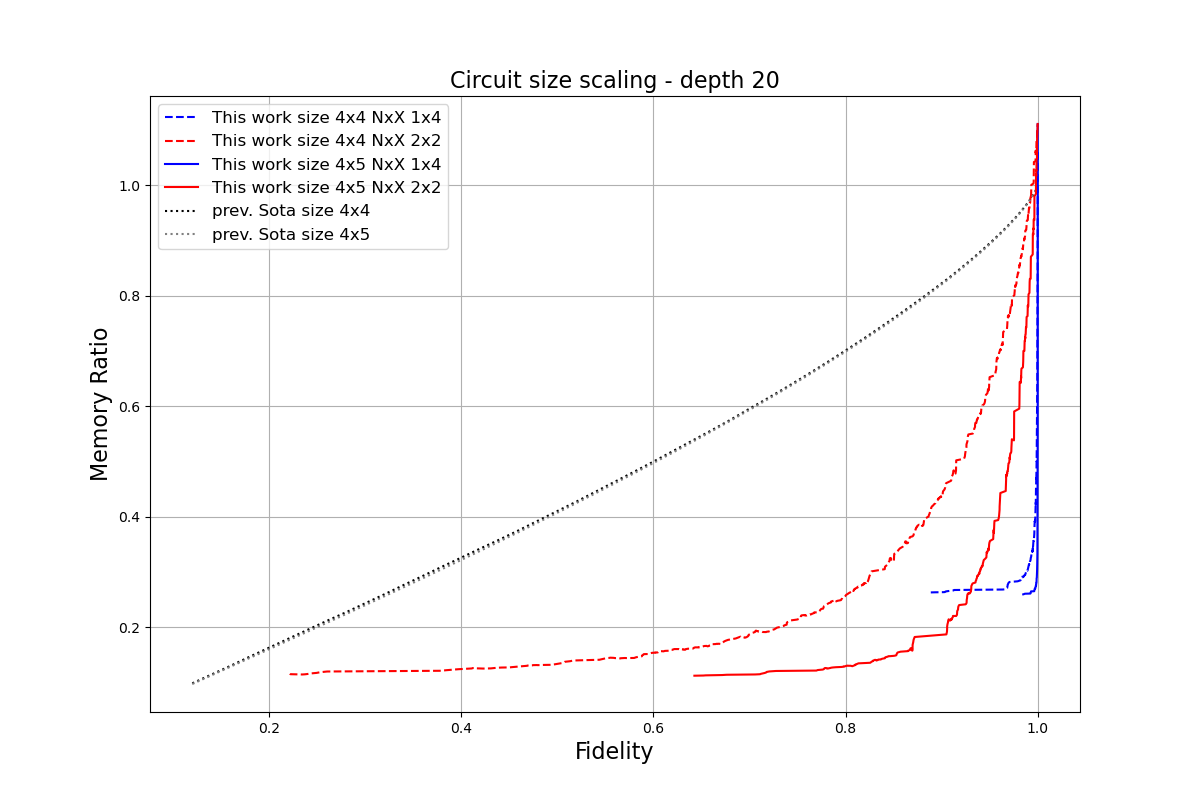}
    \caption{depth 20}
    \label{fig:lsh_fidmem_4x5}
    \end{subfigure}
    \caption{Memory ratio vs. fidelity when increasing circuit size. For the previous Sota, the trade-off between memory and fidelity stays roughly the same. But for the node replacement approach in this work, the approximation performance improves with increasing circuit size.}
    \label{fig:lsh_size}
\end{figure}

As shown in Fig. \ref{fig:lsh_size}, when the circuit size is scaled up, the previous Sota approach based on node removal doesn't change much, because its approximation performance depends on the distribution of contribution, which doesn't change much while scaling up the circuit size. 

On the other hand, the approximation performance of node replacement approach in this work increases while scaling up the circuit size. The reason is, in the same vector space, when there are more nodes, it becomes easier to find more similar nodes. Because of this, we assume the approximation performance would continue to increase as we simulate larger circuits. But simulating larger circuits were not possible with available compute resources.

\subsection{Scaling up Circuit Depth}\label{subsec:res_depth}

\begin{figure}
    
    \begin{subfigure}{1\linewidth}
    \centering\includegraphics[width=1\linewidth]{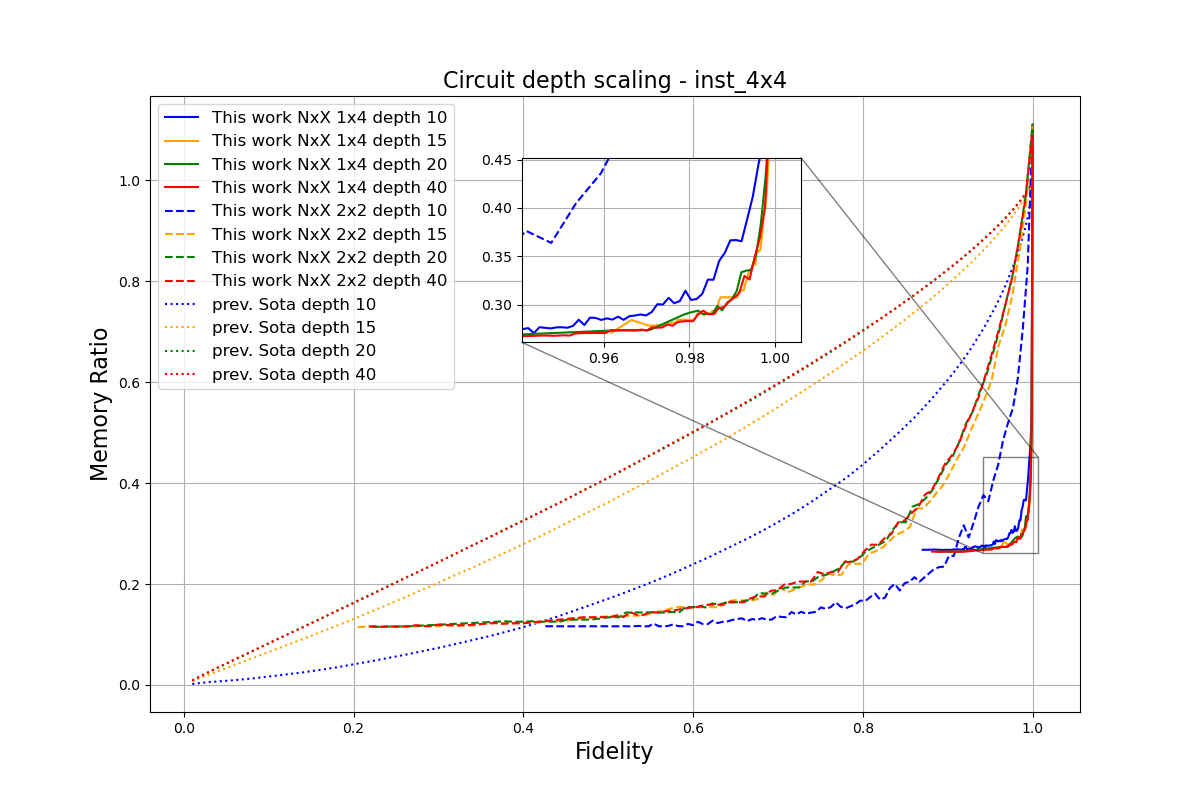}
    \caption{4x4}
    \label{fig:lsh_fidmem_4x4}
    \end{subfigure}
    \begin{subfigure}{1\linewidth}
    \centering\includegraphics[width=1\linewidth]{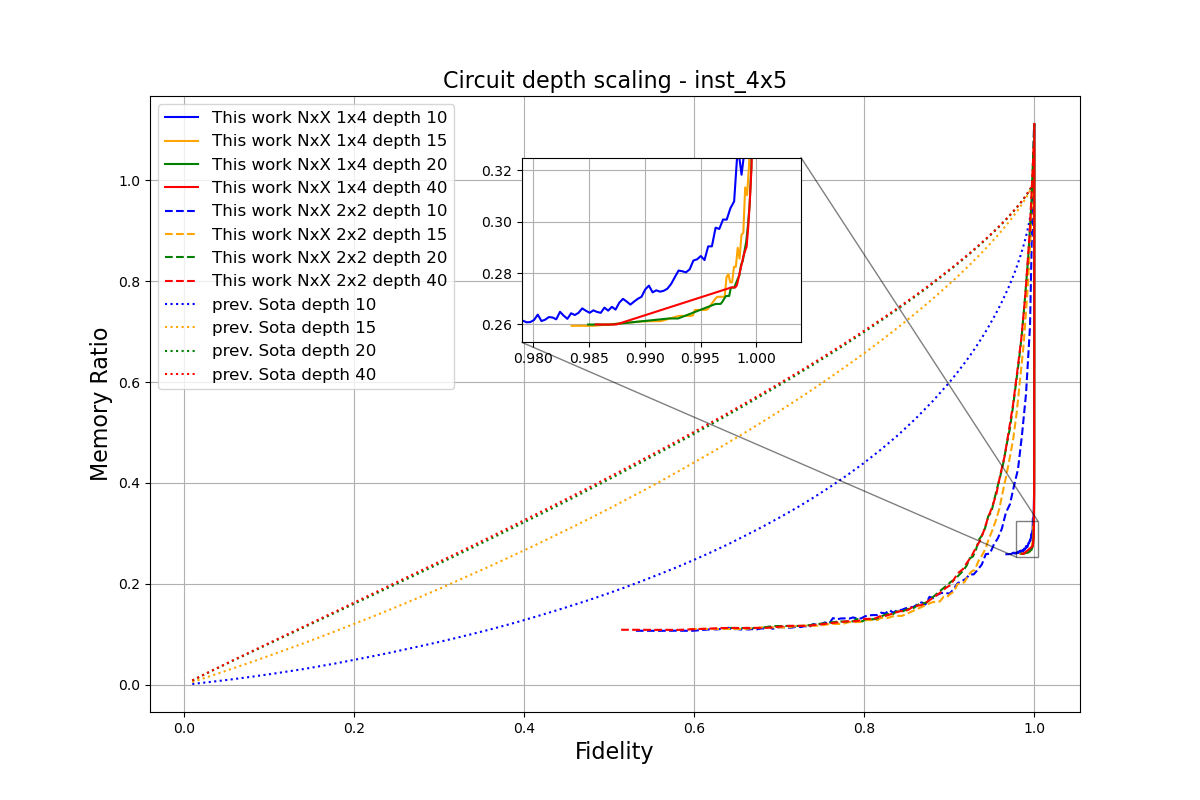}
    \caption{4x5}
    \label{fig:lsh_fidmem_4x5}
    \end{subfigure}
    \caption{Memory ratio vs. fidelity when increasing circuit depth. For the previous Sota, the trade-off between memory and fidelity becomes more linear, when the circuit depth increases. But for the node replacement approach in this work, the approximation performance degrades moderately (2x2) or even slightly increases (1x4).}
    \label{fig:lsh_depth}
\end{figure}

As shown in Fig. \ref{fig:lsh_depth}, when the circuit depth is increased, the memory-fidelity curve of previous Sota based on node removal becomes more linear, whereas the performance of this work based on node replacement decreases moderately (for 2x2) or even slightly increases (for 1x4). 

In the quantum supremacy benchmark, when the circuit depth is increased, the randomization increases, making the distribution of norm contribution of the nodes more flat, which means, there are fewer nodes with very small contribution, but instead, all nodes tend to have the same contribution. Therefore, removing each node leads to roughly the same fidelity loss. Therefore, when the circuit depth increases, as a result of the randomization, for previous Sota, the curve becomes more linear. 

On the other hand, for the node replacement approach in this work, the fidelity loss is not only related to the contribution, but also the similarity between the replaced node and replacement node (see "\nameref{subsec:fid_loss}"). As long as a very similar node for replacement can be found, the fidelity loss can be very low. Therefore, a strong better-than-linear trade-off between memory and fidelity can be achieved even with high circuit depth. 

With increasing circuit depth, the approximation performance of 1x4 and 2x2 changes in different directions, presumably due to different factors affecting the approximation performance. For 1x4, the approximation performance becomes slightly better with increasing depth. This could be caused by randomization. When most of the nodes of a level are replaced, the rest of the replacement nodes need to be very well randomized in the vector space so that each replaced node can find a similar node. But for 2x2, the approximation performance degrades moderately. The difference between 2x2 and 1x4 is, for 2x2, the sub vector of the node has higher dimensions. So it becomes more difficult to find similar nodes, since the vector space becomes sparser. When the circuit depth is low, more sub vectors can be found in local clusters in the vector space, so more similar nodes for replacement can be found.

Overall, the node replacement approach in this work shows good scaling performance regarding circuit depth and size.

%% file: sections/discussion.tex
In this work, we demonstrate using the node replacement approach for approximate simulation of quantum circuits based on decision diagrams. To test the performance, we focused on the quantum supremacy benchmark \cite{Boixo2018}, since it represents the worst case. Compared to tensor network, which efficiently represents the quantum circuit, but still needs exponential resources to compute the full state vector, decision diagrams try to efficiently represent the full state vector. While only linear trade-off between memory and accuracy has been demonstrated for tensor network, better-than-linear trade-off has been shown in previous work based on decision diagrams \cite{hillmichApprox2022}. In this work, we show substantial improvement of memory-accuracy trade-off compared to previous work, developed 4 node replacement strategies suitable for different fidelity regions, developed an LSH-based algorithm to accelerate the node replacement process, and showed very good scaling properties in terms of circuit size and depth. In particular, for the first time, we demonstrate a strong better-than-linear trade-off between memory and accuracy for the quantum supremacy benchmark at high circuit depths. 

One limitation is, the node replacement in this work is done after the DD is constructed. For DD based quantum simulations, the maximum memory requirement occurs during the construction of the DD. Therefore, this work is not able to increase the number of qubits that can be simulated within the same resources. Future work would address the integration of node replacement into the DD constrution, which has already been shown to be possible in the previous work \cite{hillmichApprox2022}. When approximation is done during DD construction, because the final fidelity is the multiplication of the fidelity at each approximation step, a fast memory reduction with a slow fidelity drop in the high fidelity region is crucial, which is already demonstrated in this work. Therefore, in this work, we show the potential of drastically reducing the memory requirement when constructing the DD of the quantum circuit.

%% file: sections/methods.tex
\subsection{Decision Diagrams for Approximate Quantum Circuit Simulation}\label{subsec:DD_intro}

\begin{figure}
    \centering\includegraphics[width=1\linewidth]{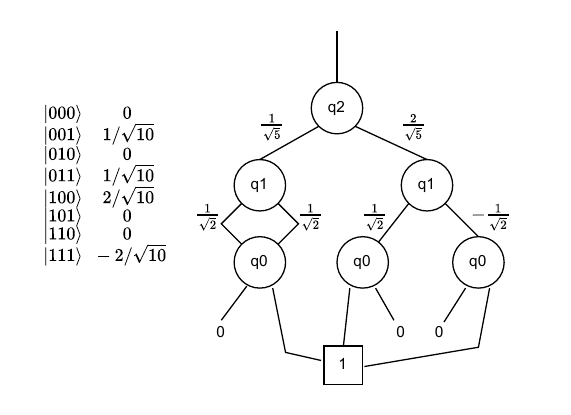}
    \caption{Quantum state with vector (left) and DD (right) representation. }
    \label{fig:dd_intro}
\end{figure}

A decision diagram (DD) represents a quantum state vector, which is a vector of amplitudes of length \(2^N\), where \(N\) is the number of qubits. As an example, the DD in Fig. \ref{fig:dd_intro} represents a state vector of 3 qubits with a state vector of 8 amplitudes. Different amplitudes in the state vector are represented by different paths from top to bottom. At the bottom of each node, choosing the left edge assigns 0 to this qubit, and choosing the right edge assigns 1 to this qubit. The amplitudes are calculated by multiplying the weights of the edges along the path. In a DD, the weight of an edge is denoted beside the edge. When no weight is denoted, the weight is 1. The weights of the two child edges of a node are normalized, so that the norm of the vector of the two weights is 1 \cite{hillmichSampling2020}. By sharing the nodes, DDs can efficiently represent a quantum state vector, as shown in the left branch of the DD in the figure. The norm contribution (or contribution) of a node is defined as the sum of squared magnitudes of amplitudes for each path passing through that node \cite{hillmichApprox2022}. For approximate simulation, in the previous work \cite{hillmichApprox2022}, for each level, the nodes are ranked based on their contribution. Then starting from the node with the smallest contribution, the contribution is accumulated, until the target fidelity loss is reached. Then these nodes are removed. 

\subsection{General Idea of Node Replacement}\label{subsec:general_idea}

The general idea of approximate simulation in this work is based on the following two ideas:

1. In the case of exact simulation, DD reduces the representation complexity of a quantum circuit by exploiting the redundancy. Specifically, it identifies \textit{identical} sub-vectors in the state vector and \textit{reuses} existing structures to represent these identical sub-vectors. 

2. Previous work on approximate simulation with DD is based on the idea that the nodes which do not significantly contribute to the overall state vector can be \textit{removed}. 

Combining these two ideas, we would like to \textit{reuse} existing nodes for nodes that do not significantly contribute to the overall state vector. This is not possible at first, because node reuse is only designed for \textit{identical} sub-vectors. And for complicated quantum circuits like the quantum supremacy benchmarks, there is almost no identical sub-vector. This conflict can be resolved by searching for nodes with \textit{similar} sub-vectors and use the node with the higher contribution to replace and approximate the removed node with a lower contribution.

\subsection{Fidelity Loss}\label{subsec:fid_loss}

\begin{figure}
    \centering\includegraphics[width=1\linewidth]{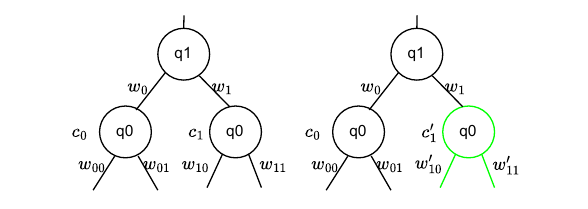}
    \caption{fid loss}
    \label{fig:dd_fid}
\end{figure}

To quantify the fidelity loss when replacing a node with another node, we consider a simple example as shown in Fig. \ref{fig:dd_fid}. Here, the right edge under node q1 is redirected to a replacement node, with a new norm contribution $c'_1$ and sub vector $[w'_{10}\; w'_{11}]$. The original state vector of the DD on the left is

\begin{equation}
|\psi\rangle = w_0w_{00}|00\rangle + w_0w_{01}|01\rangle + w_1w_{10}|10\rangle + w_1w_{11}|11\rangle
\end{equation}

The new state vector of the DD on the right is

\begin{equation}
|\phi\rangle = w_0w_{00}|00\rangle + w_0w_{01}|01\rangle + w_1w'_{10}|10\rangle + w_1w'_{11}|11\rangle
\end{equation}

The fidelity of $|\psi\rangle$ and $|\phi\rangle$ is

\begin{align}
F(|\psi\rangle, |\phi\rangle) &= |\langle \psi | \phi \rangle|^2 \nonumber \\
                      &= | (w_0w_{00})(w_0w_{00})^* + (w_0w_{01})(w_0w_{01})^* + \nonumber \\
                      &\;(w_1w_{10})(w_1w'_{10})^* + (w_1w_{11})(w_1w'_{11})^*  |^2
\end{align}

With $A=(w_0w_{00})(w_0w_{00})^* + (w_0w_{01})(w_0w_{01})^*$, we have

\begin{equation}
F(|\psi\rangle, |\phi\rangle) = |A + (w_1w_{10})(w_1w'_{10})^* + (w_1w_{11})(w_1w'_{11})^*  |^2   
\end{equation}

Without loss of generality, the norm contribution of the right node in level 0 of the original DD is $c_1=w_1w_1^*$. So we have

\begin{equation}
F(|\psi\rangle, |\phi\rangle) = |A + c_1(w_{10}w'^*_{10} + w_{11}w'^*_{11})  |^2   
\label{eq:fid1}
\end{equation}

As we know, the fidelity of the original DD with itself is 1:

\begin{align}
F(|\psi\rangle, |\psi\rangle) &= |\langle \psi | \psi \rangle|^2 \nonumber \\
                      &= | (w_0w_{00})(w_0w_{00})^* + (w_0w_{01})(w_0w_{01})^* + \nonumber \\
                      &\;(w_1w_{10})(w_1w_{10})^* + (w_1w_{11})(w_1w_{11})^*  |^2 \nonumber\\
                      &=| A + c_1(w_{10}w^*_{10} + w_{11}w^*_{11})  |^2 \nonumber\\
                      &=|1|^2\nonumber\\
                      &=1 \label{eq:fid2}
\end{align}

Since the sub vector of a node in DD is normalized \cite{hillmichSampling2020}, i.e. $w_{10}w^*_{10} + w_{11}w^*_{11}=1$, with equations \ref{eq:fid1} and \ref{eq:fid2}, we define the fidelity loss of the node replacement in this example as

\begin{align}
    L_{fid} &= c_1((w_{10}w^*_{10} + w_{11}w^*_{11})-(w_{10}w'^*_{10} + w_{11}w'^*_{11})) \nonumber \\ 
    &= c_1(1-(w_{10}w'^*_{10} + w_{11}w'^*_{11}))    
\end{align}

so that 

\begin{equation}
F(|\psi\rangle, |\phi\rangle) =|1 - L_{fid} |^2
\end{equation}

In general, when replacing a node with another node, where the sub vectors are $v_i$ and $v'_i$, and norm contributions $c_i$ and $c'_i$, respectively, the fidelity loss of the replacement can be defined as 

\begin{align}
    L_{fid,i} &= c_i(1-(v_iv'^*_i))    
\end{align}

and it can be proved that when replacing $n$ nodes, the fidelity is

\begin{equation}
F(|\psi\rangle, |\phi\rangle) =|1 - \sum_i^n L_{fid,i} |^2
\end{equation}

When \textit{removing} instead of \textit{replacing} a node, it is equivalent to replacing a node where $v'_i$ is filled with $0$. In this case, the fidelity loss is 

\begin{align}
    L_{fid,i} &= c_i(1-(v_iv'^*_i))    \nonumber \\
              &= c_i
\end{align}

By replacing the node, for each replacement, the fidelity loss is reduced by $c_i(v_iv'^*_i)$. If a very similar node can be found, $v_iv'^*_i$ is close to $1$ and $L_{fid,i}$ is close to $0$. 

In conclusion, the fidelity loss of node removal only depends on the norm contribution. But for node replacement, in addition to the norm contribution, the fidelity loss also depends on the similarity of the two nodes, and finding a similar node can reduce the fidelity loss.

\subsection{Details of Single Level Node Replacement}\label{subsec:single_level}

Based on the general idea, strategies are developed for node replacement.

We begin with the most simple case where we only replace nodes on the lowest level. 
Specifically, the node replacement is done in three steps: 

1. Rank: Similar to \cite{hillmichApprox2022}, the nodes on the lowest level are ranked according to their contribution. Then they are divided into two groups: the less important nodes, i.e. the nodes to be replaced, are kept in the 'replaced node list', and the more important nodes, i.e. the remaining nodes, are kept in the 'replacement node list'. The number of nodes in the two lists determines the trade-off between memory and accuracy, i.e. by increasing nodes in the replaced node list and reducing nodes in the replacement node list, the accuracy reduces, but more memory can be saved. 

2. Compare: In order to find the similar nodes for replacement, each node in the replaced node list is compared with each node in the replacement node list by comparing the inner product of the corresponding sub-vectors. The replacement node is the node with the largest real part of the inner product. 

\begin{figure}
    \centering\includegraphics[width=1\linewidth]{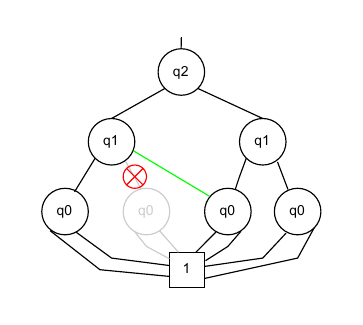}
    \caption{Single level node replacement. The red cross indicates that the edge and the node below it are deleted (replaced). The green line indicates a new edge connecting to the replacement node. (For simplicity, the weights of the edges are not shown.)}
    \label{fig:dd_1x1}
\end{figure}

3. Replace: Once a replacement node from the replacement node list is found for each node in the replaced node list, the replacement is done by redirecting the parent edge of the replaced node to the replacement node (Fig. \ref{fig:dd_1x1}).

Single level node replacement replaces nodes without further overhead (compared to the following strategies). As a result, it can achieve very high fidelity at the same memory. But it only replaces nodes on the lowest level of the DD, which limits the maximum number of nodes that can be replaced. In order to replace more nodes, we would like to consider replacing nodes on multiple levels.

\subsection{Details of Multiple Level Node Replacement}\label{subsec:multiple_level}

\begin{figure}
    \centering\includegraphics[width=1\linewidth]{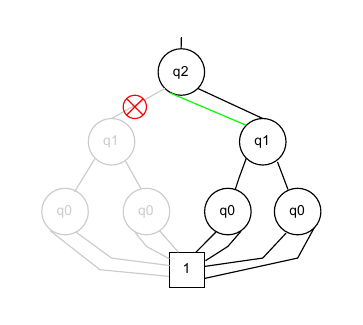}
    \caption{Multiple level node replacement. The red cross indicates that the edges and the nodes below it are deleted (replaced). The green line indicates a new edge connecting to the replacement node. Note that in this case \textit{all} edges and nodes below the red cross on the left side are replaced together by the branch on the right side. (For simplicity, the weights of the edges are not shown.)}
    \label{fig:dd_Nx1}
\end{figure}

The most straight-forward way to replace nodes on multiple levels is to rank, compare and replace nodes on a higher level instead of on the lowest level (Fig. \ref{fig:dd_Nx1}):

1. Rank: In step 1 in Section \ref{subsec:single_level}, choose a higher level and rank the nodes on the chosen level.

2. Compare: In step 2 in Section \ref{subsec:single_level}, compare nodes on the chosen level. Since the node is now on a higher level, the sub-vector is larger, because the length of the sub-vector doubles on each level, starting from the lowest level with a sub-vector of 2 entries. 

3. Replace: The same as step 3 in Section \ref{subsec:single_level}, the parent edge of the replaced node is redirected to the node found in step 2. Note that by redirecting on a higher level, all child nodes of the replaced node are replaced.

As shown in Fig. \ref{fig:dd_Nx1}, for multiple level node replacement, when a node is replaced, all its child nodes are replaced like a bundle, which causes non-optimal fidelity loss, because these child nodes might not be the best nodes to replace if they were ranked on their own level.

\subsection{Details of Independent Node Replacement}\label{subsec:iterative}

To alleviate the "bundling" problem in Section \ref{subsec:multiple_level}, and to improve the flexibility when choosing the replacement nodes, it would be desirable to rank, compare and replace the nodes on each level independently. But this comes with another issue: if a parent node is replaced, the replacement node won't be able to find the child nodes of the replaced node. 

\begin{figure}
    \centering\includegraphics[width=1\linewidth]{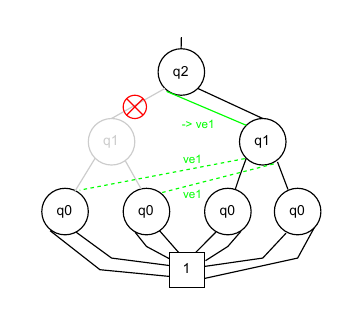}
    \caption{Independent node replacement. A node in the left branch on level 1 is replaced by a node on the right branch (red cross and green solid line). In order to go back to the child nodes of the replaced node on the left branch, virtual edges are attached to the replacement node (green dashed lines). Beside the green solid line, '\texttt{->}ve1' means 'go to virtual edge number 1'. Beside the green dashed lines, 've1', means 'virtual edge number 1'. }
    \label{fig:dd_1xXa}
\end{figure}

This issue can be resolved by introducing "virtual edges" into DD. The rank and compare steps are the same as before, but in the replace step, a virtual edge is attached to the replacement node which includes the information of the child nodes of the replaced node (Fig. \ref{fig:dd_1xXa}). This allows independent replacement on each level, but creates a memory overhead. Specifically:

1. When the child nodes of the replaced node are terminal nodes, i.e. the replaced node is on level 0, no virtual edge is attached to the replacement node, since terminal nodes are the same.

2. Starting from level 1, virtual edges is attached to the replacement node including the information of the child nodes of the replaced node. Since many replaced nodes could have the same replacement node, a replacement node could have many virtual edges. Therefore, when redirecting the parent edge of the replaced node, a "virtual edge number" is added to the parent edge, so that the child nodes of the replaced node can be found by taking the right virtual edges (Fig. \ref{fig:dd_1xXa}). 

\begin{figure*}
    \centering\includegraphics[width=1\linewidth]{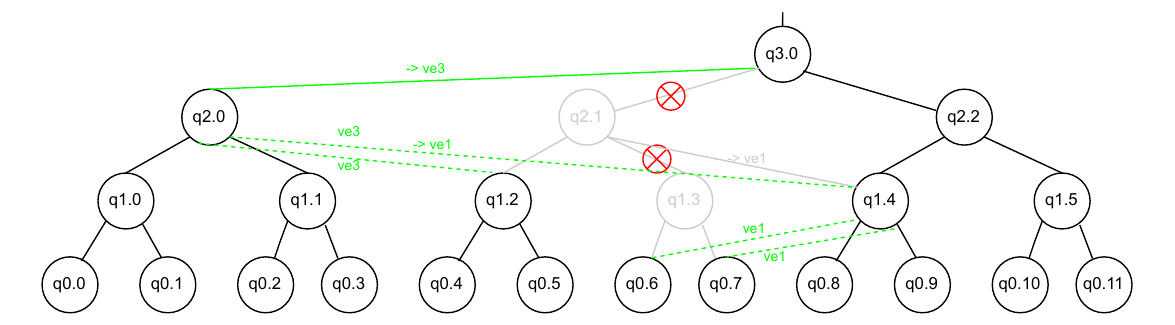}
    \caption{Independent node replacement. In this example, first, when replacing nodes on level 1,  the node q1.3 is replaced by q1.4. Virtual edges are attached to q1.4, and the edge connecting q2.1 to q1.4 contains the virtual edge number (\texttt{->}ve1). Then, when replacing nodes on level 2, q2.1 itself is replaced by q2.0, and virtual edges are attached to q2.0. To make sure the virtual edge number information of q2.1 (\texttt{->}ve1) is not lost, this information is added to the virtual edge of q2.0.}
    \label{fig:dd_1xXb}
\end{figure*}

3. When replacing a node on level 2 or above, it could happen that the child edges of a replaced node include virtual edge numbers. In this case, these virtual edge numbers are added to the virtual edge of the replacement node (Fig. \ref{fig:dd_1xXb}). This is not done for node replacement on level 1 in order to reduce memory overhead. 

In terms of memory, assume a node originally requires memory size $M$. To accommodate the virtual edge and virtual edge number, the memory size of each node (including its two child edges) is increased by $0.11M$. On level 1, each node replacement adds a virtual edge with 

\begin{equation}
(V \cdot 2 + 2)/18 \cdot M,     
\label{eq:vc1}
\end{equation}

where $V$ is the number of child nodes, which is 2 in this case. So a virtual edge adds $0.33M$ memory overhead. On level 2 and above each node replacement adds a virtual edge with 

\begin{equation}
(V \cdot 4 + 2)/18 \cdot M.     
\label{eq:vc2}
\end{equation}

With $V=2$, a virtual edge adds $0.55M$ memory overhead.

For the quantum supremacy circuit, we assume level 0 has 50\% of the nodes, level 1 has 25\% of the nodes, and so on. Based on equations \ref{eq:vc1} and \ref{eq:vc2}, in the extreme case, suppose all nodes are replaced, it would result in $25\% \cdot 0.33M + 12.5\% \cdot 0.55M + 6.25\% \cdot 0.55M + ... = 22\%M$ overhead.

In conclusion, independent node replacement improves the flexibility in choosing the replacement nodes, at the cost of some memory overhead.

\subsection{Details of Independent Multiple Level Node Replacement}\label{subsec:multi_iterative}

\begin{figure*}
    \centering\includegraphics[width=1\linewidth]{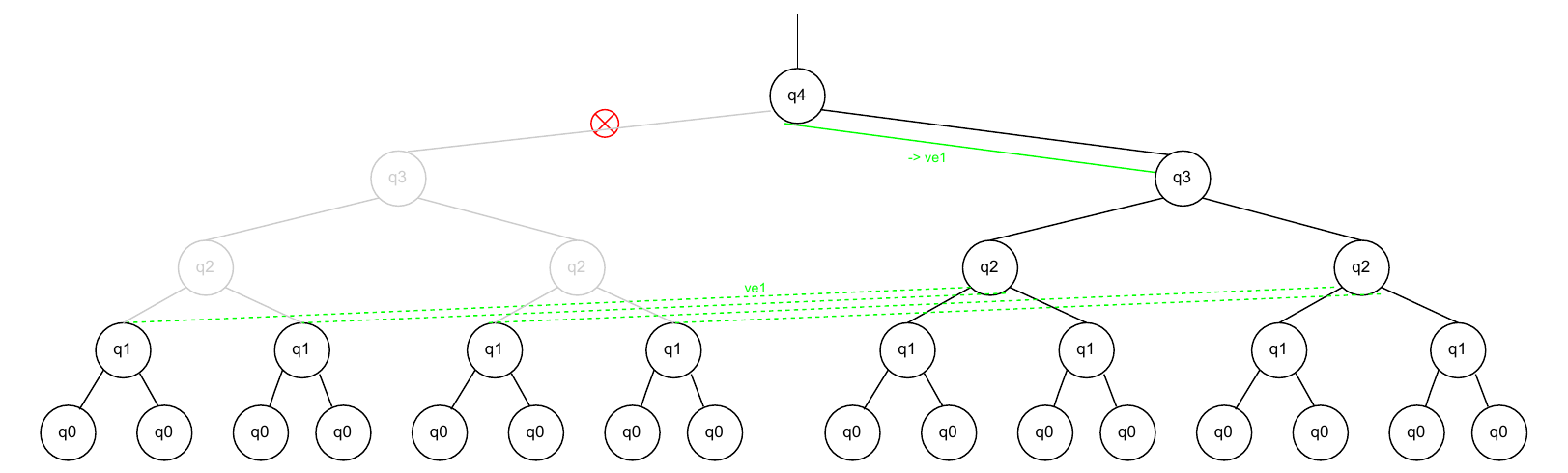}
    \caption{Independent multiple level node replacement. In this example, nodes of 2 levels (level 2 and 3) are replaced together. But level 2 and 3 are replaced independently from level 0 and 1.}
    \label{fig:dd_NxX}
\end{figure*}

Multiple level node replacement achieves the lowest memory bound in the low fidelity region. Independent node replacement alleviates the fidelity drop of multiple level node replacement in the high fidelity region, at the cost of memory overhead. A hybrid solution would be combining independent node replacement and multiple level node replacement: instead of rank, compare and replace nodes on each level, do this once every $N$ levels for $X$ times ($N$x$X$). The resulting independent multiple level node replacement would provide a solution between independent node replacement and multiple level node replacement (Fig. \ref{fig:dd_NxX}). 

 According to equations \ref{eq:vc1} and \ref{eq:vc2}, with $V=4$ ($N=2$) in this case, each virtual edge on level 3 adds $0.55M$ memory overhead, and each virtual edge on level 5 and above adds $M$ memory overhead. In the extreme case, suppose all nodes are replaced, it would result in $ 0.5^4 \cdot 0.55M + 0.5^6 \cdot M + 0.5^8 \cdot M + ... = 5.5\%M$ overhead.

%% file: sections/fid.tex
In the previous section, in order to find a replacement node, a replaced node needs to be compared with each node in the replacement node list. The exhaustive searching results in a quadratic scaling of computational complexity. To resolve this issue, we propose using Locality-Sensitive-Hashing (LSH) to reduce the computational complexity.

\subsection{General Idea of LSH}\label{subsec:lsh_idea}

The idea of node replacement is to replace nodes with similar nodes. In machine learning, approximate nearest neighbor (ANN) retrieval accomplishes a similar task, with applications like internet search engines. The idea is to calculate an index for each data vector in the data base based on a chosen distance metric. When there is a new data vector, its index can be calculated in the same way, and data vectors with the same (or similar) index in the data base can be retrieved. One algorithm family in this field is LSH. 

In this work, first we determine a threshold for norm contribution after ranking the nodes, which is used to split the nodes into the replaced node list and the replacement node list. Then we use LSH to hash each node into a bucket. Within each bucket, the nodes above the threshold are used to replace nodes below the threshold, by comparing the similarity of each pair of nodes. This is similar to the method described in the previous section, but the difference is, the replacement only happens within the same bucket. This is based on the assumption that the nodes from the same bucket are more similar than nodes from different buckets. By restricting the comparison to the most similar nodes, the computational complexity can be drastically reduced.

\subsection{Suitability of Cosine Similarity}\label{subsec:cosine_sim}

Various distance metrics are available for LSH. For the comparison of two sub vectors of quantum states in DD, cosine similarity is chosen, based on the intuition that the sub vectors in DD are normalized. In addition, the cosine similarity corresponds to the real part of the inner product of the two quantum states, which we explain in the following.

The fidelity of two quantum states $|\psi\rangle$ and $|\phi\rangle$ is defined as the square of the inner product. 

\begin{equation}
F(|\psi\rangle, |\phi\rangle) = |\langle \psi | \phi \rangle|^2    
\end{equation}

with 

\begin{equation}
\langle \psi | \phi \rangle = \sum_j \psi_j^* \phi_j
\end{equation}

where $\psi_j$ and $\phi_j$ are the $j$-th element of $\psi$ and $\phi$, respectively. Let $\psi_j = \psi_{jr} + i\psi_{ji}$ and $\phi_j = \phi_{jr} + i\phi_{ji}$, we get

\begin{equation}
\langle \psi | \phi \rangle = \sum_{j} \left( \psi_{jr} \phi_{jr} + \psi_{ji} \phi_{ji} \right) + i \sum_{j} \left( \psi_{jr} \phi_{ji} - \psi_{ji} \phi_{jr} \right)
\end{equation}

where $\sum_{j} \left( \psi_{jr} \phi_{jr} + \psi_{ji} \phi_{ji} \right)$ can be interpreted as the cosine similarity, given that the real and imaginary parts of the complex values in the state vectors are shown individually:

\begin{equation}
|\mathbf{\psi}\rangle = \begin{pmatrix} \psi_{1r} & \psi_{1i} & \psi_{2r} & \psi_{2i} & \cdots & \psi_{nr} & \psi_{ni} \end{pmatrix}
\end{equation}

\begin{equation}
|\mathbf{\phi}\rangle = \begin{pmatrix} \phi_{1r} & \phi_{1i} & \phi_{2r} & \phi_{2i} & \cdots & \phi_{nr} & \phi_{ni} \end{pmatrix}.
\end{equation}

Therefore, by looking for replacement nodes with high cosine similarity, we essentially find sub vectors with large real part of the inner product.

In the case where the two quantum states are identical, the inner product is a real number, since the imaginary parts cancel out each other due to the conjugation. In the experiments, we find out that by maximizing the real part of the inner product, the imaginary part becomes negligibly small. Therefore, maximizing the real part of the inner product leads to high fidelity of the two quantum states. 

\subsection{Super-Bit LSH}\label{subsec:sblsh}

After establishing cosine similarity as the similarity metric, we adopt the Super-Bit LSH \cite{NIPS2012_072b030b} which uses orthogonalized random projection vectors to achieve better performance compared to the widely adopted Sign-random-projection LSH \cite{10.1145/509907.509965}. The Super-Bit LSH hashes vectors of length $d$ into buckets denoted with binary codes of length $K$. In particular, the $K$ hash vectors are grouped into $lsh_l$ batches, with each batch containing $lsh_n$ hash vectors, i.e. $K = lsh_l \times lsh_n$, where the vectors within a batch are orthogonalized. In this work, $d$ corresponds to the length of the sub vector of the node in the DD. $lsh_n$ is chosen to be 2, which corresponds to the real and imaginary parts of an entry in the state vector. $lsh_l$ is a parameter which controls the number of buckets.  

\subsection{Hierarchical LSH}\label{subsec:hlsh}

The number of nodes in different levels of DD and in different benchmarks can be very different. If the parameter $lsh_l$ is too large, it would result in too many buckets, so that in each bucket, there are not enough nodes to ensure that each node can find a good replacement node. In addition, it could happen that many buckets only have one node, so that replacement wouldn't be possible. On the other hand, if $lsh_l$ is too small, there are too few buckets, and in each bucket there could be too many nodes, so that the run time is not effectively reduced. 

To avoid manually picking a value for $lsh_l$ for each level of the DD and for each benchmark, we propose using hierarchical LSH for balancing the number of nodes in the buckets. Assume there are $N$ nodes in a level. Based on the intuition that the maximum possible number of buckets is $2^{lsh_l \times lsh_n}$, and $lsh_n=2$ in this work, we start with $lsh_l = (\log_2 N) / 2$. If the nodes in a bucket is more than $\sqrt{N}$, the nodes in this buckets are hashed with additional random projection vectors. In this way, LSH works adaptively for different levels with different number of nodes.

%% file: DDApprox.bbl
\begin{thebibliography}{10}
\providecommand{\url}[1]{#1}
\csname url@samestyle\endcsname
\providecommand{\newblock}{\relax}
\providecommand{\bibinfo}[2]{#2}
\providecommand{\BIBentrySTDinterwordspacing}{\spaceskip=0pt\relax}
\providecommand{\BIBentryALTinterwordstretchfactor}{4}
\providecommand{\BIBentryALTinterwordspacing}{\spaceskip=\fontdimen2\font plus
\BIBentryALTinterwordstretchfactor\fontdimen3\font minus
  \fontdimen4\font\relax}
\providecommand{\BIBforeignlanguage}[2]{{%
\expandafter\ifx\csname l@#1\endcsname\relax
\typeout{** WARNING: IEEEtran.bst: No hyphenation pattern has been}%
\typeout{** loaded for the language `#1'. Using the pattern for}%
\typeout{** the default language instead.}%
\else
\language=\csname l@#1\endcsname
\fi
#2}}
\providecommand{\BIBdecl}{\relax}
\BIBdecl

\bibitem{grover}
\BIBentryALTinterwordspacing
L.~K. Grover, ``A fast quantum mechanical algorithm for database search,'' in
  \emph{Proceedings of the Twenty-Eighth Annual ACM Symposium on Theory of
  Computing}, ser. STOC '96.\hskip 1em plus 0.5em minus 0.4em\relax New York,
  NY, USA: Association for Computing Machinery, 1996, p. 212-219. [Online].
  Available: \url{https://doi.org/10.1145/237814.237866}
\BIBentrySTDinterwordspacing

\bibitem{shor}
P.~Shor, ``Algorithms for quantum computation: discrete logarithms and
  factoring,'' in \emph{Proceedings 35th Annual Symposium on Foundations of
  Computer Science}, 1994, pp. 124--134.

\bibitem{Nielsen_Chuang_2010}
M.~A. Nielsen and I.~L. Chuang, \emph{Quantum Computation and Quantum
  Information: 10th Anniversary Edition}.\hskip 1em plus 0.5em minus
  0.4em\relax Cambridge University Press, 2010.

\bibitem{Harrow2017}
\BIBentryALTinterwordspacing
A.~W. Harrow and A.~Montanaro, ``Quantum computational supremacy,''
  \emph{Nature}, vol. 549, no. 7671, pp. 203--209, Sep 2017. [Online].
  Available: \url{https://doi.org/10.1038/nature23458}
\BIBentrySTDinterwordspacing

\bibitem{Boixo2018}
\BIBentryALTinterwordspacing
S.~Boixo, S.~V. Isakov, V.~N. Smelyanskiy, R.~Babbush, N.~Ding, Z.~Jiang, M.~J.
  Bremner, J.~M. Martinis, and H.~Neven, ``Characterizing quantum supremacy in
  near-term devices,'' \emph{Nature Physics}, vol.~14, no.~6, pp. 595--600, Jun
  2018. [Online]. Available: \url{https://doi.org/10.1038/s41567-018-0124-x}
\BIBentrySTDinterwordspacing

\bibitem{qflex}
\BIBentryALTinterwordspacing
B.~Villalonga, S.~Boixo, B.~Nelson, C.~Henze, E.~Rieffel, R.~Biswas, and
  S.~Mandr{\`a}, ``A flexible high-performance simulator for verifying and
  benchmarking quantum circuits implemented on real hardware,'' \emph{npj
  Quantum Information}, vol.~5, no.~1, p.~86, Oct 2019. [Online]. Available:
  \url{https://doi.org/10.1038/s41534-019-0196-1}
\BIBentrySTDinterwordspacing

\bibitem{markov2008}
\BIBentryALTinterwordspacing
I.~L. Markov and Y.~Shi, ``Simulating quantum computation by contracting tensor
  networks,'' \emph{SIAM Journal on Computing}, vol.~38, no.~3, pp. 963--981,
  2008. [Online]. Available: \url{https://doi.org/10.1137/050644756}
\BIBentrySTDinterwordspacing

\bibitem{niemann2016}
P.~Niemann, R.~Wille, D.~M. Miller, M.~A. Thornton, and R.~Drechsler, ``Qmdds:
  Efficient quantum function representation and manipulation,'' \emph{IEEE
  Transactions on Computer-Aided Design of Integrated Circuits and Systems},
  vol.~35, no.~1, pp. 86--99, 2016.

\bibitem{zulehner2019}
A.~Zulehner and R.~Wille, ``Advanced simulation of quantum computations,''
  \emph{IEEE Transactions on Computer-Aided Design of Integrated Circuits and
  Systems}, vol.~38, no.~5, pp. 848--859, 2019.

\bibitem{boixo2018b}
\BIBentryALTinterwordspacing
I.~Markov, A.~Fatima, S.~Isakov, and S.~Boixo, ``Quantum supremacy is both
  closer and farther than it appears,'' \emph{arxiv (not yet submitted)}, 2018.
  [Online]. Available: \url{https://arxiv.org/abs/1807.10749}
\BIBentrySTDinterwordspacing

\bibitem{hillmichApprox2022}
\BIBentryALTinterwordspacing
S.~Hillmich, A.~Zulehner, R.~Kueng, I.~L. Markov, and R.~Wille, ``Approximating
  decision diagrams for quantum circuit simulation,'' \emph{ACM Transactions on
  Quantum Computing}, vol.~3, no.~4, jul 2022. [Online]. Available:
  \url{https://doi.org/10.1145/3530776}
\BIBentrySTDinterwordspacing

\bibitem{hillmichSampling2020}
S.~Hillmich, I.~L. Markov, and R.~Wille, ``Just like the real thing: fast weak
  simulation of quantum computation,'' in \emph{Proceedings of the 57th
  ACM/EDAC/IEEE Design Automation Conference}, ser. DAC '20.\hskip 1em plus
  0.5em minus 0.4em\relax IEEE Press, 2020.

\bibitem{NIPS2012_072b030b}
\BIBentryALTinterwordspacing
J.~Ji, J.~Li, S.~Yan, B.~Zhang, and Q.~Tian, ``Super-bit locality-sensitive
  hashing,'' in \emph{Advances in Neural Information Processing Systems},
  F.~Pereira, C.~Burges, L.~Bottou, and K.~Weinberger, Eds., vol.~25.\hskip 1em
  plus 0.5em minus 0.4em\relax Curran Associates, Inc., 2012. [Online].
  Available:
  \url{https://proceedings.neurips.cc/paper_files/paper/2012/file/072b030ba126b2f4b2374f342be9ed44-Paper.pdf}
\BIBentrySTDinterwordspacing

\bibitem{10.1145/509907.509965}
\BIBentryALTinterwordspacing
M.~S. Charikar, ``Similarity estimation techniques from rounding algorithms,''
  in \emph{Proceedings of the Thiry-Fourth Annual ACM Symposium on Theory of
  Computing}, ser. STOC '02.\hskip 1em plus 0.5em minus 0.4em\relax New York,
  NY, USA: Association for Computing Machinery, 2002, p. 380-388. [Online].
  Available: \url{https://doi.org/10.1145/509907.509965}
\BIBentrySTDinterwordspacing

\end{thebibliography}
